# Interface agents: A review of the field

*Stuart E. Middleton*
Intelligence, Agents and Multimedia group (IAM group)

University of Southampton

Southampton, SO17 1BJ, UK.

Email: sem99r@ecs.soton.ac.uk Web: http://www.ecs.soton.ac.uk/~sem99r

## Abstract

This paper reviews the origins of interface agents, discusses challenges that exist within the interface agent field and presents a survey of current attempts to find solutions to these challenges. A history of agent systems from their birth in the 1960's to the current day is described, along with the issues they try to address. A taxonomy of interface agent systems is presented, and today's agent systems categorized accordingly. Lastly, an analysis of the machine learning and user modelling techniques used by today's agents is presented.

### Keywords

Agents, interface agents, survey, review

## Table of Contents

## Table of Figures

# 1 Introduction

The 1990's have seen the dawn of a new paradigm in computing - software agents. Many researchers are currently active in this vibrant area, drawing from more traditional research within the artificial intelligence (AI) and human computer interaction (HCI) communities. Kay [24] and others argue that one aspect of software agent systems, the interface agent, has the potential to revolutionize computing as we know it, allowing us to advance from direct manipulation of systems to indirect interaction with agents. Removing the requirement for people to manage the small details of a task liberates individuals, empowering them to accomplish goals otherwise requiring experts.

Now, this may be the future of computing but software agents originated from the field of artificial intelligence, back in the 1950's. The next section describes some of the important landmarks that happened along the way to where we are today.

## 2 History of software agents

Alan Turing, famous for his work on computability [76], posed the question "Can machines think?" [77]. His test, where a person communicates via a Teletype with either a person or a computer, became known as the Turing test. The Turing test requires a conversational computer to be capable of fooling a human at the other end. It is the Turing test that inspired the birth of the artificial intelligence community.

The discipline of artificial intelligence (AI) was born in the 1950's. Marvin Minsky, after some work with neural networks (deemed a failure at the time due to the difficulty of learning weights), teamed up with John McCarthy at MIT to work on symbolic search-based systems. At the same time at Carnegie-Mellon, Allen Newell and Herbert Simon were successfully exploring heuristic search to prove logic theorems. Initial successes thus led to heuristic search of symbolic representations becoming the dominant approach to AI.

The 1960's saw much progress. Now at Stanford, McCarthy [38] had just invented LISP and set about representing the world with symbols, using logic to solve problems [39]. At the same time Newell [50] created the General Problem Solver which, given a suitable representation, could solve any problem. Problems solved were in simple, noise and error free symbolic worlds, with the assumption that such solutions would generalize to allow larger, real world problems to be tackled. Researchers did not worry about keeping computation on a human time-scale, using the increases in hardware performance to constantly increase the possible search space size, thus solving increasingly impressive problems.

During the 1970's, search became well understood [51]. Symbolic systems still dominated, with continuing hardware improvements allowing steady, successful progress. Robots were created, for example Shakey [52], that lived in special block worlds, and could navigate around and stack blocks sensibly. Such simplified worlds avoided the complexity of real world problems. The assumption, underpinning all the symbolic research, that simple symbolic worlds would generalize to the real world, was about to be found wanting.

In the 1980's, expert systems were created to try to solve real problems. McCarthy [40] had realized that "common sense" was required in addition to specialized domain knowledge to solve anything but simple microworld problems. A sub-field of AI, knowledge representation, came into being to examine approaches to representing the everyday world. Unfortunately the idea of "common sense" proved impossible to represent, and knowledge-based systems were widely viewed to have failed to solve real-world problems. At the same time, the backpropagation algorithm [67] caused a resurgence of interest in connectionist approaches, previously deemed a failure, and Minsky [42] examined an agent-based approach for intelligence.

The late 1980's and early 1990's saw the decline of search-based symbolic approaches. Brooks [10] convincingly challenged the basic assumptions of the symbolic approaches, and instead created embodied, grounded systems for robots using the "world as its own best model". This bottom up approach was termed *nouvelle AI*, and had some initial successes. However, it too failed to scale up to real-world problems of any significant complexity. Connectionist approaches were aided by new parallel hardware in the early 1990's, but the complexity of a parallel architecture led such systems to fail in the marketplace.

Knowledge engineering, now widely seen as costly and hard to re-use, was superseded by machine learning techniques borrowed from AI. Towards the end of the 1990's, pattern-learning algorithms [44] could classify suitable domains of knowledge, such as news stories and examination papers, with as much accuracy as manual classification. Hybrids of traditional and nouvelle AI started to appear as new approaches were sought.

The mid 1990's saw Negroponte [49] and Kay's [24] dream of indirect HCI coupled with Minsky's [42] ideas on intelligence lead to the new field of agent-based computing. Experiments with interface agents that learnt about their user [33], and multi-agent systems where simple agents interacted to achieve their goals [80] dominated the research. Such agent systems were all grounded in the real world, using proven AI techniques to achieve concrete results (applying the maxim "a little AI goes a long way").

User modelling changed in the 1990's too, moving from the static hand crafted representations of the 1980's to dynamic behaviour based models [25]. Machine learning techniques proved particularly adept at identifying patterns in user behaviour.

# 3 Issues and challenges for interface agents

Maes [33] describes interface agents as follows:

> "Instead of user-initiated interaction via commands and/or direct manipulation, the user is engaged in a co-operative process in which human and computer agents both initiate communication, monitor events and perform tasks. The metaphor used is that of a *personal assistant* who is *collaborating with the user* in the same work environment."

The motivating concept behind Maes' interface agents is to allow the user to delegate mundane and tedious tasks to an agent assistant. Her own agents follow this direction, scheduling and rescheduling meetings, filtering emails, filtering news and selecting good books. Her goal is to reduce the workload of users by creating personalized agents to which personal work can be delegated.

There are many interface agent systems and prototypes, inspired by Maes early work, situated within a variety of domains. The majority of these systems are reviewed and categorized in the next section. Common to these systems, however, are three issues that must be addressed before successful user collaboration with an agent can occur:

- Knowing the user
- Interacting with the user
- Competence in helping the user

Knowing the user involves learning user preferences and work habits. If an assistant is to offer help at the right time, and of the right sort, then it must learn how the user prefers to work. An eager assistant, always interrupting with irrelevant information, would just annoy the user and increase the overall workload.

The following challenges exist for systems trying to learn about users:

- Extracting the users' goals and intentions from observations and feedback
- Getting sufficient context in which to set the users' goals
- Adapting to the user's changing objectives
- Reducing the initial training time

At any given time, an interface agent must have an idea of what the user is trying to achieve in order to be able to offer effective assistance. In addition to knowing what the user's intentions are, there must be sufficient contextual information about the user's current situation to avoid irrelevant agent help. Machine learning techniques help here, but which should be used and why?

Another problem is that regular users will typically have numerous concurrent tasks to perform. If an agent is to be helpful with more than one task, it must be able to discover when the user has stopped working on one job, and is progressing to another – but what is the best way to detect this?

Users are generally unwilling to invest much time and effort in training software systems. They want results early, before committing too much to a tool. This means that interface agents must limit the initial period during which the agent learns enough about the user to offer useful help. What impact does this have on an agent's learning ability?

A metaphor for indirect HCI has yet to reach maturity, so remains an open question. Lessons have been learned from direct manipulation interfaces. Users need to feel in control, expectations should not be unduly inflated and user mistakes should not be penalized [53].

Interacting with the user thus presents the following challenges:

- Deciding how much control to delegate to the agent
- Building trust in the agent
- Choosing a metaphor for agent interaction
- Making simple systems that novices can use

It is known from direct manipulation interfaces that users want to feel in control of what their tools are doing. By the nature of an autonomous interface agent, some control has been delegated to it, in order for it to do its task. The question is, how do we build the users' trust, and once a level of trust is established how much control do we give to the agents? Shneiderman [74] argues for a combination of direct manipulation and indirect HCI, promoting user understanding of agents and the ability for users to control agent behaviour directly. How can we use these guiding principles in our systems?

Interface metaphors, such as the desktop metaphor, guide users in the formation of useful conceptual models a system. New metaphors will be required for indirect HCI, presenting agents in a way helpful to users new to the system. Ideally, interface agents should be so simple to use that delegating tasks becomes a natural way of working, amenable to the novice user – but what is a natural way of working with agents?

Lastly, there is the issue of competence. Once the agent knows what the user is doing and has a good interaction style, it must still formulate a plan of action that helps, not hinders, the user. The challenges are:

- Knowing when (and if) to interrupt the user
- Performing tasks autonomously in the way preferred by the user
- Finding strategies for partial automation of tasks

There is very little current research into how users can be best helped. Work from other disciplines, such as computer supported co-operative working (CSCW), can help but real user trials are needed to demonstrate and evaluate effectiveness and usefulness of the personalized services performed by interface agents [54]. If an agent does not reduce the workload of a real user in a real work setting, it is less than useful.

# 4 Taxonomy of interface agent systems

Several authors [54] [80] have suggested taxonomies for software agents as a whole, but they tend to address interface agents as a monolithic class, citing a few examples of the various prototypical systems. With the maturing of the agent field, and the growing number of interface agents reported in the literature, a more detailed analysis is warranted. Mladenić [46] goes some way to achieving this requirement, adopting a machine learning view of interface agents.

Interface agents can be classified according to the role they perform, technology they use or domain they inhabit. Interface agents are moving from research to commercial exploitation, significantly increasing the roles and domains for agents as entrepreneurs find new ways to exploit new markets. The fundamental technology behind the agents, however, is undergoing less radical change, and thus provides a more stable basis on which to build a useful taxonomy.

On this basis a survey of current interface agent technology has been performed. The next section details the actual agent systems and prototypes reviewed. The result is a non-exclusive taxonomy of the technologies that specific agent systems support.

- Character-based agents
- Social agents
    - Recommender systems
- Agents that learn about the user
    - Monitor user behaviour
    - Receive user feedback
        - Explicit feedback
        - Initial training set
    - Programmed by user
- Agents with user models
    - Behavioural model
    - Knowledge-based model
    - Stereotypes

Character-based agents employ advanced “character” based interfaces, representing real world characters (such as a pet dog or a human assistant [34]). Such agents draw on existing real-world protocols, already known to even novice users, to facilitate more natural interaction. There are also applications in the entertainment domain, creating state of the art virtual worlds populated by believable agents.

Social agents talk to other agents (typically other interface agents of the same type) in order to share information. This technique is often used to bootstrap new, inexperienced interface agents with the experience of older interface agents (attached to other users).

Recommender systems are a specific type of social agent. They are also referred to as collaborative filters [60], finding relevant items based on the recommendations of others. Typically, the user's own ratings are used to find similar users, with the aim of sharing recommendations on common areas of interest.

Agents employing a learning technology are classified according to the type of information required by the learning technique and the way the user model is represented. Algorithms requiring an explicit training set employ supervised learning, while those without a training set use unsupervised learning techniques [44]. There are three general ways to learn about the user: monitor the user, ask for feedback or allow explicit programming by the user.

Monitoring the user's behaviour produces unlabelled data, suitable for unsupervised learning techniques. This is generally the hardest way to learn, but is also the least intrusive. If the monitored behaviour is assumed to be an example of what the user wants, a positive example can be inferred.

Asking the user for feedback, be it on a case-by-case basis or via an initial training set, produces labelled training data. Supervised learning techniques can thus be employed, which usually outperform unsupervised learning. The disadvantage is that feedback must be provided, requiring an investment of effort (often significant) in the agent by the user.

User programming involves the user changing the agent explicitly. Programming can be performed in a variety of ways, from complex programming languages to the specification of simple cause/effect graphs. Explicit programming requires significant effort by the user.

User modelling [25] comes in two varieties, behavioural and knowledge-based. Knowledge-based user modelling is typically the result of questionnaires and studies of users, hand-crafted into a set of heuristics. Behavioural models are generally the result of monitoring the user during an activity. Stereotypes [64] can be applied to both cases, classifying the users into groups (or stereotypes), with the aim of applying generalizations to people in those groups.

Specific interface agents will often implement several of the above types of technology, and so would appear in multiple classes. A common example is an agent that learns about the user and also supports a user model. The presented taxonomy ought to be robust to the increase in new systems, since the fundamental technology of machine learning and user modelling are unlikely to change as quickly.

# 5 Review of current interface agent systems and prototypes

A comparison of interface agents is difficult since there are no widely used standards for reporting results. Where machine learning techniques are employed, standard tests such as precision and recall provide useful metrics for comparing learning algorithms. However, the best test of an interface agent's ability to help a user is a user trial. Unfortunately, user trials in the literature do not follow a consistent methodology.

The analysis in this paper will focus on classifying the agent, identifying techniques used such as specific machine learning techniques or user modelling types, and where applicable results published by the original author. Comparisons of systems can thus be made on a qualitative basis.

## 5.1 Review of current agent systems

What follows is a review of known interface agent systems. Agents are examined by application domain, so that similar types of interface agents can be compared. The machine learning algorithms specified here are described later in the glossary.

### 5.1.1 Auction/market domain

*Kasbah* [36] is a market system in which each user has an agent. The user programs the agent with a buying behaviour profile, and the agent negotiates to buy and sell items for the user.

Results: Users wanted more "human like" negotiation from the agents, otherwise well received.

*Sardine* [47] is an auction agent that tries to purchase an airline ticket for the user, based on some specified preferences. The user's agent negotiates with travel agents to secure the best deal.

### 5.1.2 Believable/entertainment domain

*ACT* [28] is an addition to the ALIVE system. It is a creature within the ALIVE world, observing the user and learning chains of actions. It tries to help the user by completing new action chains in the pattern of previous ones.

*ALIVE* [35] is a "magic mirror" system to a 3D world. Interactive agents (such as a dog) exist for users to play with. Gesture recognition, and competing goal architecture is employed.

*Cathexis* [37] is a believable agent with modelled emotions, as is the *Oz* project [4].

### 5.1.3 Email filtering domain

*MailCat* [72] filters email by providing a choice of folders to the user. TF-IDF vectors are created for existing emails, and cosine similarity used to match new emails. The user has the final say, choosing one of the suggested folders or moving messages manually.

Results: 0.3 second classification time, 60-80% accuracy giving user one choice, 80-98% accuracy giving user 3 choices of folder.

*MAGI* [17] filters emails, monitoring user behaviour and receiving relevance feedback. CN2 and IBPL are used to classify emails.

*Maxims* [33] filters email by learning repetitive actions the user performs. It monitors user actions using memory-based reasoning to discover patterns. Agents can share expertise with other agents, and user programming is allowed.

*Re:Agent* [9] is an email filter that accepts user provided keywords for its groupings. TF-IDF vectors are created for each email, along with the TF of the user provided keywords. This representation is then clustered using a nearest neighbour and neural network clustering algorithm (for comparison).

Results: classification accuracy – neural network 94.8 ± 4.2%, nearest neighbour 96.9 ± 2.3%; high accuracy due to simple classification task (into "work" or "other" categories).

### 5.1.4 Expert assistance domain

*Coach* [73] is a LISP help system that monitors user mistakes and offers unsolicited advice. A knowledge-based user model is supported, with the concept of user experience stereotypically represented. Heuristics adjust the model based on user mistakes.

Results: Student performance improved, knowledge of functions improved by a factor of 5.

*Eager* [13] automates observed repetitive HyperCard actions. It monitors the user looking for behaviour patterns, and creates helpful macros from them.

Results: Users felt a loss of control; macros for some irrelevant small patterns were created.

*GALOIS* [68] monitors the use of an application, and offers expert advise when users are lost or being inefficient. An initial knowledge-based user profile is constructed from personal information, then a behavioural model built by observing user actions. Stereotypes are used to classify users, thus allowing customized help.

*GESIA* [11] helps expert system developers by suggesting predicted actions. A Bayesian network models user behaviour, allowing predictions with the help of hard-coded domain knowledge.

*Open Sesame!* [20] observes user actions and offers to automate repetitive tasks. The ART-2 learning algorithm is used.

Results: Only 2/129 suggestions were followed – system deemed to have failed; action patterns do not generalize across situations well.

### 5.1.5 Matchmaking domain

*ExpertFinder* [78] monitors users' Java code and finds people who use the same classes. TF-IDF vectors represent code files, and cosine similarity is used to find similar people.

*ReferralWeb* [23] builds a social network from publicly available web pages. People's names are extracted from pages, and co-occurrence of names within pages imply a social connection. Queries such as "list docs close to Mitchell" can thus be issued.

*Yenta* [16] allows user agents to "find" each other, and determine commonality of interests. The SMART algorithm initially classifies user emails, newsgroups and created files in order to build an interest profile. Agents then find each other in the Yenta system, compare profiles for similarity, and suggest other agents to try.

Results: Halves the worst-case search space, robust to removal of agents.

### 5.1.6 Meeting schedulers

*CAP* [43] is a calendar manager, monitoring email and scheduling software to detect meeting patterns. Decision trees (ID3), using information gain to select features, are converted to production rules.

Results: 31-60% accuracy (average of 47%) not sufficient for automation, rules were human readable which improved user understanding.

*Meeting scheduling agent* [32] schedules meetings by learning repetitive actions the user performs. Memory-based reasoning and reinforcements learning are used. Users can give explicit feedback.

Results: Confidence for correct predictions settles at 0.8 to 1.0. Confidence for incorrect predictions settles at 0 to 0.2. Some rouge confidence values remain after settling time.

Haynes' [19] meeting scheduler assigns an agent to each user, who then programs it with their personal preferences. The agents then negotiate meeting times with each other.

### 5.1.7 News filtering domain

*ANATAGONOMY* [69] is based on the Krakatoa Chronicle, providing a personalized newspaper. Implicit feedback from user activity has been added.

Results: 1-10% error after 3 days settling time.

*Butterfly* [29] finds interesting conversations within Usenet newsgroups. The user initially provides keywords, and term frequency similarity between newsgroups and the user's profile is computed.

*IAN* [17] filters Usenet news, taking relevance feedback from the user. C4.5 rule induction with TF keyword selection (low entropy words being removed) is compared to IBPL (as used in MAGI).

Results: accuracy – C4.5 broad topics 70%, narrow topics 25-30% IBPL broad topics 59-65%, narrow topics 40-45%.

The *Krakatoa Chronicle* [22] is a personalized newspaper which adapts to its users' preferences. User reading is monitored and relevance feedback accepted. The SMART algorithm is used, with TF-IDF, to represent articles and compute similarity.

*NewsDude* [6] reads interesting news articles via a speech interface. The news source is Yahoo! News, with an initial training set of interesting news articles provided by the user. Length of listening time provides implicit user feedback on articles read out. A short-term user model is based on TF-IDF (cosine similarity), and long-term model based on a naïve Bayes classifier (multi-variate Bernoulli formulation).

Results: Accuracy 60-76% (using hybrid of long and short term models), F1 measure 25-60%.

*NewsWeeder* [27] filters Usenet newsgroups, taking user relevance feedback. It uses a "bag of words" approach, with stemming. TF-IDF and minimum description length are compared, using cosine similarity.

Results: TF-IDF precision was 37-45%, MDL precision was 44-59% (best).

*NewT* [33] filters articles from Usenet Netnews. NewT uses a vector space model to represent news articles. An initial training set and user relevance feedback trains the filter and user programming is allowed.

Results: Users liked the system and found it useful; the simple keyword representation was a limitation.

Pannu's [56] learning personal agent finds relevant information from Usenet news (e.g. conference proceedings). The user, along with feedback on examples suggested, provides an initial training set. TF-IDF and a 3 layer backpropagation neural network were compared.

Results: neural network precision 94% recall 60%, TF-IDF precision 93% recall 100% (best).

### 5.1.8 Recommender systems

*GroupLens* [26] recommends newsgroup articles based on other's ratings. Users are asked to rate every article they read, and Pearson correlation coefficient-based prediction is used to find similar users. A new article will be rated according to the ratings of similar users.

Results: Early users receive little reward for rating articles, lazy users do not rate articles and new users do not find similar users until they rate many articles.

*PHOAKS* [75] monitors newsgroup articles, and extracts web link recommendations by a set of heuristics. A set of collective recommendations for topics is thus compiled.

Results: 88% precision, 87% recall.

*Ringo* [33] recommends music, based on other people's recommendations. Uses collaborative recommendation. A Pearson r correlation algorithm is used to determine similarity.

Results: Initial start-up a problem, users relying just on recommendations miss new material, which will not get rated until someone chooses it.

*Siteseer* [66] recommends web sites based on the overlap of bookmark groups with other users. Users with similar bookmark groups are found, and unseen web sites recommended from their groups.

Results: New users receive no help as no similarity can be found with others, average of 18% confidence in recommendation for topics with 15-20 bookmarks.

### 5.1.9 Web domain

*AARON* [17] is the same as LAW, but AutoClass is used as a learning algorithm.

*Amalthaea* [48] observes user browsing behaviour and assists the user in finding interesting WWW information. Browser history, bookmarks and other agent profiles initialise the system. Relevance feedback is recorded and document representation is by stemmed, keyword vectors. A genetic algorithm is used.

Results: After 5000 user feedback instances, error averaged 7% with a large scatter (0 – 30%). Sudden changes in user interest were tracked after about 20 generations.

*ARACHNID* [41] is a spider that crawls a digital library or the web, starting from the users' bookmarks, searching for user provided keywords. The user provides feedback on the pages found. A genetic algorithm is used in addition to reinforcement learning.

Results: Average search length (number of irrelevant docs before a relevant one is found) shorter than breath-first search by a factor of 10. More sophisticated techniques not compared.

*CiteSeer* [8] helps users perform keyword searches for publications by using citations in documents. Heuristics extract citations, titles and abstracts, then algorithms (TF-IDF, LikeIt) classify publications based on stemmed words. Citation links lead to new search keywords.

*Do-I-Care* [58] is an agent that monitors known sites, and reports any interesting changes. Mutual information selects terms, and a Bayesian classifier determines the similarity of changes to known examples of relevant changes. The user gives feedback on relevance of notifications (by email) and other people's agents can share examples of relevant changes.

*Fab* [2] is a recommendation system for web pages. Relevance feedback is obtained from users, so similar users' recommendations can be used. In addition, content-based analysis of documents finds new documents of interest.

Results: ndpm measure (of profile accuracy) 0.2-0.4

*Jasper* [14] finds relevant information within a limited WWW library. The user provides interest groups and keywords and gives relevance feedback. User behaviour, such as reading and authoring, is also monitored. Documents and keywords are clustered using a hierarchical agglomerative process, with features extracted from keywords and metadata (e.g. URL, title).

*LAW* [17] is an agent that finds interesting web pages for its user. The agent monitors user browsing and asks for relevance feedback. TF, TF-IDF and term relevance are compared for feature selection algorithms, along with IBPL and C4.5 for learning algorithms.

Results: accuracy TF 60-80%(best), TF-IDF 55-70%, term relevance 60-65% and accuracy IBPL 65-83%(best), C4.5 55-65%.

*Letizia* [28] suggests interesting web links, based on monitored user browsing. TF similarity between documents is used, and heuristics infer user preferences. For example, a short view time implies that the user regards the page as irrelevant.

*Let's browse* [31] helps users of TV systems collaboratively browse the web. Based on Letizia, TF-IDF scans of users' home pages provide an initial profile, and group browsing creates trials of web pages. Browsing behaviour is used to infer page relevance.

Results: 50 (as opposed to 10 in Letizia) keywords needs, reflecting a groups wider interests; system well received by users (no controlled experiments however).

*LIRA* [1] finds interesting web pages via a heuristic search and presents them in a daily newspaper, upon which the users provide relevance feedback. TF-IDF, on stemmed words, with cosine similarity is used.

Results: LIRA matched human performance; pages were very similar to each other.

*Margin Notes* [62] adds a suggestions list to the side of the web browser. The user provides an initial list of interesting documents, which are converted to vectors by the Savant algorithm (similar to TF-IDF). The current web page provides the context for suggestions, with the top suggestion being displayed (summary and a link).

Results: only 6% of suggestions were followed; users found suggestion summaries useful even without following links.

*Personal WebWatcher* [45] observes user browsing, and suggests interesting web pages. A "bag of words" representation is used, selecting features based on mutual information. Naive Bayes and nearest neighbour algorithms are compared.

Results: classification accuracy Bayes 83-95%, nearest neighbour 90-95% (best).

*SAIRE* [55] is a large multi-agent system helping users search NASA's web resources. Knowledge-based, stereotypical user modelling is employed along with semantic networks, relevance rankings and similarity based (keyword and topic) classification.

*Syskill and Webert* [57] rates web links for relevance based on a user profile. An initial training set and explicit relevance feedback is provided. Simple naïve Bayes, nearest neighbour, decision tree (ID3) and TF-IDF algorithms are compared.

Results: Average precision ratings are TF-IDF 85%, Bayes 80%, nearest neighbour 80%, ID3 73%. Nearest neighbour is thought to be best overall (being more consistent than TF-IDF), especially if many examples are available.

*WebACE* [18] monitors the user's browsing, and suggests interesting new pages. Browsed pages are classified (via clustering), and search queries generated. New pages found that are similar to the users' browsed pages are suggested to the user. Principle component divisive partitioning (PCDP - document vectors split by TF into a binary tree) and an association rule discovery method are compared to Autoclass and hierarchical agglomeration clustering.

Results: Speed to find new low entropy pages, Autoclass 38 mins, HAC 100 mins, PCDP and association rule < 2 mins (best).

*WebMate* [12] monitors web browsing and compiles a newspaper of interesting pages. Users provide positive feedback while browsing and relevance feedback. TF-IDF vectors used with cosine similarity measures.

Results: average accuracy 52% for top 10 recommendations, 30.4% for all recommendations; Accuracy lowered by web advertisements and irrelevant text surrounding articles.

*WebWatcher* [21] is a tour guide for a web site. Pages are represented using TF-IDF term vectors. User browsing behaviour is monitored, and reinforcement learning used to find the best path to pages related to user supplied keywords.

Results: TF-IDF accuracy 43%, Human accuracy 48%.

*WBI* [3] monitors user browsing, offering simple but helpful services. Keyword analysis classifies web pages, and interest predictions are offered on links to pages. It is commercially available from IBM.

Pazzani's [59] adaptive web site agent suggests similar documents to read for web browsers of a particular site. It uses publication references, download frequency and TF-IDF keyword vectors (with cosine similarity measure) to suggest other documents of interest.

Results: 68% increase in publications downloaded (technical papers domain), 16% increase (goat domain).

### 5.1.10 Other domains

*CILA* [7] is an agent tested in an artificial, abstract domain. It tests constructive induction-based learning against AQ15c and selective induction. User monitoring, relevance feedback, initial training sets and social collaboration with other agents are supported (in its abstract world).

Results: Constructive induction was most accurate but only on an artificial domain.

*CIMA* [28] is a text prediction agent, which suggests completions of sentences in a text editor. Heuristics learn from observed examples, hints and partial specifications.

*COLLAGEN* [65] is an agent whose interaction style is modelled on human collaboration. Agents and users share a goal and plan, and communicate actions and results via a dialogue.

*Grammex* [30] learns grammar (e.g. email structure) from user examples and performs actions when it detects new occurrences of this grammar. The user programs by example, using a direct manipulation interface to teach the agent.

*Mondrian* [28] learns graphical operations which are explicitly programmed by users. Novices can then use these operations.

Waszkiewicz [79] builds a personal travel assistant, aiming to meet the FIPA 1997 travel scenario. The user specifies preferences and a travel agent suggests a flight. Case-based reasoning (a case retrieval net) is used to build a user profile from past requests. User confirmation is sought before booking.

*Softbot* [15] plans internet-based actions from incomplete user goal specifications (e.g. "send mail to Mitchell"). A planning library of schemata is used, written by hand in Prolog.

*Remembrance agent* [63] suggests documents related to the user's current context. Emails and on-line documents are monitored, and the SMART algorithm used to match context and documents.

Results: Email most useful for up-to-date contextual information, RA preferred over a search engine or Margin notes [61].

*UCI GrantLearner* [5] is a system to identify interesting grant funding opportunities. It uses the same user model learning techniques as Syskill & Webert.

Schlimmer [70] describes a text-completion agent, using finite state machines with embedded decision trees to predict user's textual input and offer a shortcut to completion.

Results: FSM compares well with ID4 and Bayes, with a hybrid of FSM and ID4 working best. Accuracy of 12-82% was seen, depending on the topic of the notes being taken.

## 5.2 Classification of agent systems

Figure 1 lists the agent systems mentioned in the previous section and shows how they are classified. The distribution of technologies within today's agents can be clearly seen.

| | Character-based agent | Social agent | Recommender system | Monitor user behaviour | Explicit feedback | Initial training set | Programmed by user | Behavioural model | Knowledge-based model | Stereotypes |
|---|---|---|---|---|---|---|---|---|---|---|
| ACT | o | | | o | | | | o | | |
| ALIVE | o | | | | | | | | | |
| Cathexis | o | | | | | | | | | |
| CAP | | o | | o | | | | o | | |
| COLLAGEN | | o | | | | | | | | |
| Do-I-Care | | o | | | o | o | | | | |
| ExpertFinder | | o | | o | | | | o | | |
| Kasbah | | o | | | | | o | | | |
| Maxims | | o | | o | | | o | o | | |
| Meeting scheduling agent (Maes) | | o | | o | o | | | o | | |
| Sardine | | o | | | | | o | | | |
| Yenta | | o | | | | o | | | | |
| GroupLens | | | o | | o | | | | | |
| PHOAKS | | | o | o | | | | | | |
| Ringo | | | o | | o | | | | | |
| Siteseer | | | o | | | o | | | | |
| Referral Web | | | o | | | o | | | | |
| Fab | | | o | o | o | | | | | |
| AARON | | | | o | o | | | o | | |
| Adaptive web site agent (Pazzani) | | | | o | | o | | o | | |
| Amalthaea | | | | o | o | | | o | | |
| ANATAGONOMY | | | | o | o | | | o | | |
| CILA | | | | o | o | o | | o | | |
| CIMA | | | | o | | | | o | | |
| Coach | | | | o | | | | | o | o |
| Eager | | | | o | | | | o | | |
| GALOIS | | | | o | | | | o | o | o |
| GESIA | | | | o | | | | o | | |
| Jasper | | | | o | o | | | o | | |
| Krakatoa Chronicle | | | | o | o | | | o | | |
| Letizia | | | | o | | | | o | | |
| LAW | | | | o | o | | | o | | |
| Let's Browse | | | | o | | o | | o | | |
| MAGI | | | | o | o | | | o | | |
| Margin Notes | | | | o | | o | | | | |
| NewsDude | | | | o | | o | | o | | |
| Open Sesame! | | | | o | | | | o | | |
| Personal WebWatcher | | | | o | | | | o | | |
| Remembrance agent | | | | o | | | | o | | |
| Sentence completer (Schlimmer) | | | | o | | | | o | | |
| Travel assistant (Waszkiewicz) | | | | o | o | | | o | | |
| WebACE | | | | o | | | | o | | |
| WebMate | | | | o | o | | | o | | |
| WebWatcher | | | | o | | o | | o | | |
| WBI | | | | o | | | | o | | |
| ARACHNID | | | | | o | o | | | | |
| IAN | | | | | o | | | | | |
| Learning personal agent (Sycara) | | | | | o | o | | | | |
| LIRA | | | | | o | | | o | | |
| NewsWeeder | | | | | o | o | | o | | |
| NewT | | | | | o | o | o | o | | |
| Syskill & Webert | | | | | o | o | | o | | |
| MailCat | | | | | o | o | | | | |
| Re:Agent | | | | | o | o | | | | |
| UCI GrantLearner | | | | | o | o | | o | | |
| Butterfly | | | | | | o | | | | |
| CiteSeer | | | | | | o | | | | |
| Grammex | | | | | | | o | | | |
| Meeting scheduler (Haynes) | | | | | | | o | | | |
| Mondrian | | | | | | | o | | | |
| Softbot | | | | | | | o | | | |
| SAIRE | | | | | | | | | o | o |

**Figure 1 Classification of agent systems**

# 6 Conclusions

Behavioural user modelling dominates the interface agent field. Behavioural user models are usually based on monitoring the user and/or asking the user for relevance feedback. The statistical information generated by these approaches is usually fed to some form of machine learning algorithm.

Almost all the non-social interface agents reviewed use a textual, content-based learning approach, deriving information from user emails, web documents, papers and other such sources. The “bag of words” document representation dominates the field, with TF-IDF proving to be a popular choice of word weighting algorithm. Relevance feedback is normally used to provide labels for documents, allowing supervised learning techniques to be employed.

Social interface agents, using collaborative learning approaches, are in the minority, but have proved useful when systems have many users. The main problem with a purely social system is that performance is initially poor until such a time as enough people are using the system. Hybrid systems, using content-based techniques to bootstrap the learning process, do address this problem to some extent and comprise about half the social agents reviewed.

Experimental results, where published, tend to be either qualitative user studies or quantitative measurements against benchmark document collections. Nawana [54] has observed that it is yet to be proven that interface agents actually help people at all. To gain evidence to this end, experiments with real users in real work settings must be performed, and ways found to compare different approaches with criteria such as helpfulness and usefulness. Only then will the interface agent community be able to put some scientific weight behind the many claims made over the last few years for “intelligent agents”.

The question is: how do current systems measure up to the challenges previously identified within the interface agent field?

*Knowing the user*

Supervised machine learning techniques require large (in the order of 100,000’s) labelled document corpuses to be effective. Since most users will not have 100,000 examples of what they like, interface agent profile accuracy falls below what most people find acceptable. Unsupervised learning techniques do allow the web’s millions of unlabelled documents to be used, but currently have poor accuracy compared with supervised learning.

Explicit user feedback allows users to label document examples, which can increase profile accuracy. Unfortunately, users are typically unwilling to commit much effort to a system unless it gives them a reward in a reasonable timeframe. This problem exists with both collaborative and content-based systems.

Current systems do learn about the user, but not to the accuracy that would allow confident delegation of full control to important tasks. Today’s interface agents can make useful suggestions, but they still need a human to ultimately check them.

*Interacting with the user*

Experiments with believable agents build realistic agents so people can interact with them in ways they feel are natural. Unfortunately, natural interaction leads to the expectation that the agent will behave in a completely natural way. When, for example, an Einstein agent appears before them they will assume they can speak to the agent as they would a human. Agent systems are not that sophisticated, so users are left disappointed or confused.

Most agent systems avoid presenting an image at all, preferring to work in the background. This is the most practical approach given today's technology, but can leave users feeling they have lost control of the "hidden" agents.

There is still much debate over the best way to interact with agents [74], with no sign of a conclusion in the near future. Only time and lots of experimental interfaces will tell how best to proceed.

*Competence in helping the user*

Most interface agent work has concentrated on learning about the user, with the assumption that once an agent knows what the user wants it can provide effective help. Planning and CSCW techniques can be utilised, but experiments are required to demonstrate competence, and show which techniques are best used in various types of situation.

# 7 Glossary of machine learning terminology

For a more detailed description of machine learning, [71] provides an excellent overview of machine learning techniques, as does [44].

***AQ15c*** – Rule learning algorithm. Rules are added until an example is fully classified, using a general to specific approach.

***ART-2*** – Adaptive resonancy theory-2, a neural network approach.

***Association rule discovery*** – Data mining technique to discover rules that associate items within a database. It is related to the induction of classification rules.

***Autoclass*** – Bayesian classifier for unsupervised classification, based on a classical mixture model.

***Backpropagation*** - Neural network algorithm for updating hidden layer weights. A reliable technique, it is the backbone of many neural networks.

***Bag of words*** – Document representation consisting of a list of words and the number of times the words appear (term frequency).

***Bayesian network*** – A probabilistic network storing the believed conditional dependencies between variables. Joint probability distributions specify the probability of a set of values to a set of variables.

***C4.5*** – ID3 variant, applying rule post-pruning and other additional techniques.

***Case retrieval net*** – Type of case-based reasoning algorithm, storing similarity between connected elements within the network.

***Case-based reasoning*** – A memory-based reasoning algorithm, using a symbolic representation of cases (as opposed to a vector space approach).

***Constructive induction-based learning*** – Inductive logic programming approach, where background knowledge is used to augment the set of predicates used.

***Cosine similarity*** – dot product measure of the distance between two vectors. This is used to measure similarity between two documents when the vector space represents document features.

***Decision tree*** – Algorithm using a tree, with each node of the tree dividing the hypothesis space using an attribute. As the tree is traversed, from top to bottom, the hypothesis space is increasingly sub-divided until only one hypothesis is left. Decision trees can be easily converted into classification rules.

***Entropy*** – A measure of the “purity” of a collection of examples. It measures the difference between the number of positive and negative examples (zero is a “pure” or perfectly balanced set).

***Finite state machine*** – Decision trees situated as states within a finite state machine (used in [70]).

***Genetic algorithm*** – Learning algorithm that supports a population of hypotheses, and evolves them by survival of the fittest, cross-over (combining two successful hypotheses) and mutation.

***Hierarchical agglomeration clustering*** – Starts with one document cluster, and agglomerates the most similar clusters until the desired number of clusters exists. TF-IDF is often used to weight document vectors.

***IBPL*** – A memory-based reasoning algorithm, storing situations and classifying by comparing similarity between situations.

***ID3*** – Classic decision tree learning algorithm. Uses information gain to select node terms.

***ID4*** – Variant on ID3.

***Information gain*** – Measure of the expected reduction in entropy of a term.

***Keyword vector*** – A vector of keywords. Vector has length equal to the number of terms in a document set, and values are the frequency of each term (usually applied to a document to give a document vector).

***LikeIt*** – An algorithm to measure distance between two strings, based on the number of character/symbolic transformations to make first string into the second string.

***Memory-based reasoning*** – Example-based classifier, storing all labelled examples in memory, and determining similarity at run-time. The nearest neighbour algorithm is an example of memory-based reasoning, where labelled examples are held as document vectors.

***Minimum description length*** – Principle that favours hypothesis with the smallest number of terms over longer ones. This ensures simple hypotheses dominate.

***Multi-variate Bernoulli formulation*** – A specific type of naïve Bayesian classifier.

***Mutual information*** - Type of information measure, used to weight terms with respect to positive examples.

***Naïve Bayes classifier*** – Probabilistic classifier based around Bayes theorem. Term probabilities are assigned to classes, and for a new document the probability of belonging to any particular class is computed.

***Nearest neighbour*** – Learning algorithm that measures the distance between document vectors within a vector space representation. The distance indicates similarity of documents (the nearest neighbours) – cosine similarity is often used.

***Neural network*** – Network of units, with inputs usually representing terms and outputs classes. Connections between units have weights, which are trained by loading examples (using a training rule such as backpropagation to update weights).

***npdm metric*** – comparative measurement of how documents rank relative to each other.

***Pearson correlation*** – Type of information measure, used to weight terms with respect to positive and negative examples.

***Principal component divisive partitioning*** – Top down clustering method, splitting the training set until small enough clusters have been formed. A binary tree thus holds the clusters.

***Reinforcement learning*** – Learning algorithm where actions produce rewards or penalties, thus the most rewarding sequence of actions is reinforced (hence learnt).

***Rocchio classifier*** – Learning algorithm, often used with TF-IDF weightings. Class term vectors are computed by summing positive example weights and subtracting negative example weights.

***Savant*** – Type of TF-IDF algorithm.

***Selective induction learning*** – this is the same as Inductive logic programming (see AQ15c algorithm).

***SMART*** – An indexing engine, which converts documents into document vectors. It uses TF-IDF weighting.

***Stemming*** – Removal of suffixes from words. Used to reduce the number of terms that are synonyms in a textual document.

***TF*** – Term frequency. The number of times a term (often a word or phrase) occurs within a document.

***TF-IDF*** – Term frequency – Inverse Document Frequency. Algorithm for assigning weights to terms in a document set, biased to weight the most discriminating terms highest.

## 8 References

[1] **Balabanović, M. Shoham, Y. Yun, Y.** "An Adaptive Agent for Automated Web Browsing", *In Journal of Visual Communication and Image Representation, 6(4), December, 1995*

[2] **Balabanović, M. Shoham, Y.** "Fab: Content-Based, Collaborative Recommendation", *Communications of the ACM 40(3), March 1997, 67-72*

[3] **Barrett, R. Maglio, P.P. Kellem, D.C.** "WBI: A Confederation of Agents that Personalize the Web", *In Autonomous Agents 97, Marina Del Rey, California USA*

[4] **Bates, J.** "The role of emotion in believable agents", *Communications of the ACM 37(7), July 1994, 122-125*

[5] **Billsus, D. Pazzani, M.** "Learning Probabilistic User Models", *In Workshop Notes of "Machine Learning for User Modeling", Sixth International Conference on User Modeling, Chia Laguna, Sardinia, 1997*

[6] **Billsus, D. Pazzani, M.J.** "A Personal News Agent that Talks, Learns and Explains", *In Autonomous Agents 98, Minneapolis MN USA*

[7] **Bloedorn, E. Wnek, J.** "Constructive Induction-based Learning Agents: An Architecture and Preliminary Experiments", *In Proceedings of the First International Workshop on Intelligent Adaptive Systems (IAS-95), Melbourne Beach, Florida, 1995, 38-51*

[8] **Bollacker, K.D. Lawrence, S. Giles, C.L.** "CiteSeer: An Autonomous Web Agent for Automatic Retrieval and Identification of Interesting Publications", *In Autonomous Agents 98, Minneapolis MN USA*

[9] **Boone, G.** "Concept Features in Re :Agent, an Intelligent Email Agent", *In Autonomous Agents 98, Minneapolis MN USA*

[10] **Brooks, R. A.** "Intelligence Without Reason", *In Proceedings of the 1991 International Joint Conference on Artificial Intelligence, 569-695*

[11] **Brown, S.M. Santos, E., Banks, S.B.** "A Dynamic Bayesian Intelligent Interface Agent", *Dept of Electrical and Computer Engineering, Air Force Institute of Technology, Wright-Patterson AFB, OH 45433-7765*

[12] **Chen, L. Sycara, K.** "WebMate: A Personal Agents for Browsing and Searching", *In Autonomous Agents 98, Minneapolis MN USA*

[13] **Cypher, A.** "Eager: Programming repetitive tasks by example", *In Human factors in computing systems conference proceedings on Reaching through technology, April 27 - May 2, 1991, New Orleans, LA USA, 33-39*

[14] **Davies, J. Weeks, R. Revett, M. McGrath, A.** "Using Clustering in a WWW Information Agent", *In 18th BCS IR Colloquium, Manchester, UK, April 1996*

[15] **Etzioni, O.** "A softbot-based interface to the internet", *Communications of the ACM 37(7), July 1994, 72-76*

[16] **Foner, L.N.** “Yenta: A Multi-Agent, Referral-Based Matchmaking System”, *In Autonomous Agents 97, Marina Del Rey, California USA*

[17] **Green, C.L. Edwards, P.** “Using Machine Learning to Enhance Software Tools for Internet Information Management”, *In AAAI-96 Workshop on Internet-Based Information Systems, WS-96-06, AAAI Press, 1996, 48-55*

[18] **Han, E. Boley, D. Gini, M. Gross, R. Hastings, K. Karypis, G. Kumar, V. Mobasher, B. Moore, J.** “WebACE : A Web Agent for Document Categorization and Exploration”, *In Autonomous Agents 98, Minneapolis MN USA*

[19] **Haynes, T. Sen, S. Arora, N. Nadella, R.** “An automated meeting scheduling system that utilizes user preferences”, *In Autonomous Agents 97, Marina Del Rey, California USA*

[20] **Hoyle, M.A. Lueg, C.** “Open Sesame !: A Look at Personal Assistants”, *Proceedings of the International Conference on the Practical Application of Intelligent Agents (PAAM97), London, 1997, 51-60*

[21] **Joachims, T. Freitag, D. Mitchell, T.** “WebWatcher: A Tour Guide for the World Wide Web”, *In Proceedings of IJCAI97, August 1997*

[22] **Kamba, T. Bharat, K. Albers, M.C.** "The Krakatoa Chronicle: An Interactive, Personalized Newspaper on the Web", *Proceedings of WWW4, Boston, USA, December 1995*

[23] **Kautz, H. Selman, B. Shah, M.** “Referral Web: Combining Social Networks and Collaborative Filtering”, *Communications of the ACM 40(3), March 1997, 63-65*

[24] **Kay, A.** “User interface: A personal view”, *In: Laurel. B. (ed.). The art of Human-Computer Interface Design, Addison-Wesley, 1990, 191-207*

[25] **Kobsa A.** “User Modeling: Recent work, prospects and Hazards”, *In Adaptive User Interfaces: Principles and Practice Schneider-Hufschmidt, M. Kühme, T. Malinowski, U. (ed) North-Holland 1993*

[26] **Konstan, J.A. Miller, B.N. Maltz, D. Herlocker, J.L. Gordon, L.R. Riedl, J.** “GroupLens: Applying Collaborative Filtering to Usenet News”, *Communications of the ACM 40(3), March 1997, 77-87*

[27] **Lang, K.** “NewsWeeder: Learning to Filter NetNews”, *In ICML95 Conference Proceedings, 1995, 331-339*

[28] **Lieberman, H. Maulsby, D.** “Instructible agents: Software that just keeps getting better”, *IBM systems journal 35(3&4), 1996*

[29] **Lieberman, H. Maes, P. Van Dyke, N.W.** “Butterfly: A Conversation-Finding Agent for Internet Relay Chat”, *Proceedings of the 1999 International Conference on Intelligent User Interfaces, January 1999*

[30] **Lieberman, H. Nardi, B.A. Wright, D.** “Training Agents to Recognize Text by Example”, *In Autonomous Agents 99, Seattle WA USA*

[31] **Lieberman, H. van Dyke, N. Vivacqua, A.** “Let’s Browse : a collaborative browsing agent”, *Knowledge-Based Systems, Vol. 12, Dec. 1999, 427-431*

browsing agent", *Knowledge-Based Systems, Vol. 12, Dec. 1999, 427-431*

[32] **Maes, P. Kozierok, R.** "A learning interface agent for scheduling meetings", *In Proceedings of ACM SIGCHI International Workshop on Intelligent User Interfaces, ACM Press, NY, 1993, 81-88*

[33] **Maes, P.** "Agents that reduce work and information overload", *Communications of the ACM 37(7) July 1994*

[34] **Maes, P.** "Articifial Life meets Entertainment: Lifelike Autonomous Agents", *Communications of the ACM 38(11), November 1995, 108-114*

[35] **Maes, P. Darrell, T. Blumberg, B. Pentland, A.** "The ALIVE System: Wireless, Full-Body Interaction with Autonomous Agents", *ACM Multimedia Systems, Special Issue on Multimedia and Multisensory Virtual Worlds, ACM Press, Spring 1996*

[36] **Maes, P. Chavez, A.** "Kasbah: An Agent Marketplace for Buying and Selling Goods", *Proceedings of the First International Conference on the Practical Application of Intelligent Agents and Multi-Agent Technology, London, UK, April 1996*

[37] **Maes, P. Velásquez, J.D.** "Cathexis: A Computational Model of Emotions", *In Autonomous Agents 97, 1997*

[38] **McCarthy, J.** "Recursive Functions of Symbolic Expressions", *CACM 3, 1960, 184-195*

[39] **McCarthy, J. Hayes, P. J.** "Some Philosophical Problems from the Standpoint of Artificial Intelligence", *Machine Intelligence 4 1969, 463-502*

[40] **McCarthy, J.** "Some expert systems need common sense", *In Computer Culture: The Scientific, Intellectual and Social Impact of the Computer, vol. 426, Heinz P. (ed.) 1994 Annals of the New York Academy of Sciences, 1983, 129-137*

[41] **Menczer, F. Belew, R.K.** "Adaptive Information Agents in Distributed Textual Environments", *In Autonomous Agents 98, Minneapolis MN USA*

[42] **Minsky, M.** "The Society of Mind", *Simon and Schuster, New York, NY 1986*

[43] **Mitchell, T.M. Caruana, R. Freitag, D. McDermott, J. Zabowski, D.** "Experience with a learning personal assistant", *Communications of the ACM 37(7), July 1994, 81-91*

[44] **Mitchell T. M.** "Machine Learning", *McGraw-Hill, 1997*

[45] **Mladenić, D.** "Personal WebWatcher: design and implementation", *Technical Report IJS-DP-7472, Department for Intelligent Systems, J. Stefan Institute*

[46] **Mladenić, D. Stefan, J.** "Text-Learning and Related Intelligent Agents: A Survey", *IEEE Intelligent Systems, 1999, 44-54*

[47] **Morris, J. Maes, P.** "Negotiating Beyond the Bid Price", *In Workshop Proceedings of the Conference on Human Factors in Computing Systems (CHI*

*2000), The Hague, The Netherlands, April 1-6, 2000*

[48] **Moukas, A.** "User modelling in a Multi-Agent Evolving System", *In International Conference on User Modelling '97, Machine Learning in User Modelling Workshop Notes, Chia Laguna, Sardinia, 1997*

[49] **Negroponte, N.** "The Architecture Machine; Towards a more Human Environment", *MIT Press, Cambridge, Mass. 1970*

[50] **Newell, A. Shaw, J. C. Simon, H.** "A General Problem-Solving Program for a Computer", *Computers and Automation 8(7), 1959, 10-16*

[51] **Nilsson, N. J.** "Problem-Solving Methods in Artificial Intelligence", *McGraw-Hill, New York, NY, 1971*

[52] **Nilsson, N. J.** "Shakey the Robot", *SRI A.I. Center Technical Note 323, April 1984*

[53] **Norman D.A.** "How might people interact with agents" *Communications of the ACM 37(7) July 1994, 68-71*

[54] **Nawana, H.** "Software agents: an overview", *In The Knowledge Engineering Review, Vol 11:3, 1996, 205-244*

[55] **Odubiyi, J.B. Kocur, D.J. Weinstein, S.M. Wakim, N. Srivastava, S. Gokey, C. Graham, J.** "SAIRE – A scalable agent-based information retrieval engine", *In Autonomous Agents 97, Marina Del Rey, California USA*

[56] **Pannu, A.S. Sycara, K.** "A Learning Personal Agent for Text Filtering and Notification", *In Proceedings of the International Conference of Knowledge Based Systems, 1996*

[57] **Pazzani, M. Muramatsu, J. Billsus, D.** "Syskill & Webert: Identifying interesting web sites", *In Proceedings of the National Conference on Artificial Intelligence, 1996*

[58] **Starr, B. Ackerman, M.S. Pazzani, M.** "Do-I-Care : A Collaborative Web Agent", *ACM CHI'96, April 1996*

[59] **Pazzani, M.J. Billsus, D.** "Adaptive Web Site Agents", *In Proceedings of the Third International Conference on Autonomous Agents (Agents '99), Seattle, Washington, 1999*

[60] **Resnick, P. Varian, H. R.** "Recommender systems", *Communications of the ACM 40(3) March 1997, 56-58*

[61] **Rhodes, B.J.** "Just-In-Time Information Retrieval", *PhD thesis, June 2000*

[62] **Rhodes, B.J.** "Margin Notes: building a contextually aware associative memory", *In IUI 2000: 2000 International Conference on Intelligent User Interfaces, New Orleans, Louisiana, January 9-12, 2000, ACM.*

[63] **Rhodes, B.J. Starner, T.** "Remembrance Agent: A continuously running automated information retrieval system", *In The proceedings of the First International Conference on The Practical Application of Intelligent Agents*

*(PAAM96), 487-495*

[64] **Rich, E.** "User modelling via Stereotypes", *Cognitive Science 3 1979, 329-354*

[65] **Rich, C. Sidner, C.L.** "COLLAGEN: When Agents Collaborate with People", *In Autonomous Agents 97, Marina Del Rey, California USA*

[66] **Rucker, J. Polanco, M.J.** "Siteseer: Personalized Navigation for the Web", *Communications of the ACM 40(3), March 1997, 73-75*

[67] **Rumelhart, D. E. Hilton, G. E. Williams, R. J.** "Learning Internal Representations by Error Propagation", *In "Parallel Distributed Processing", Rumelhart D. E. and McClelland J. L., MIT Press, Camberidge, MA, 1986*

[68] **Sagula, J.E. Puricelli, M.F. Bobeff, G.J. Martin, G.M. Carlos, E.P.** "GALOIS: An Expert-Assistant Model", *In Autonomous Agents 97, Marina Del Rey, California USA*

[69] **Sakagami, H. Kamba, T. Koseki, Y.** "Learning personal preferences on online newspaper articles from user behaviors", *In 6th International World Wide Web Conference, 1997, pp. 291–300*

[70] **Schlimmer, J.C. Hermens, L.A.** "Software agents: Completing Patterns and Constructing User Interfaces", *Journal of Artificial Intelligence Research 1 (1993), 61-89*

[71] **Sebastiani, F.** "Machine Learning in Automated Text Categorisation", *Consiglio Nazionale delle Ricerche, Via S. Maria, 46-56126 Italy*

[72] **Segal, R.B. Kephart, J.O.** "MailCat: An Intelligent Assistant for Organizing E-Mail", *In Autonomous Agents 99, Seattle WA USA*

[73] **Selker, T.** "Coach: A Teaching Agent that Learns", *Communications of the ACM 37(7), July 1994, 92-99*

[74] **Shneiderman, B. Maes, P.** "Direct manipulation vs interface agents", *Interactions: new visions of human-computer interaction Nov-Dec 1997*

[75] **Terveen, L. Hill, W. Amento, B. McDonald, D. Crester, J.** "PHOAKS: A System for Sharing Recommendations", *Communications of the ACM 40(3), March 1997, 59-62*

[76] **Turing, A. M.** "On Computable numbers with an application to the Entscheidungsproblem", *Proc. London Math. Soc. 42, 1937, 230-65*

[77] **Turing A. M.** "Computing Machinery and Intelligence", *Mind 59, Oct 1950, 433-460*

[78] **Vivacqua, A.** "Agents for Expertise Location", *In Autonomous Agents 98, Minneapolis MN USA*

[79] **Waszkiewicz, P. Cunningham, P. Byrne, C.** "Case-based User Profiling in a Personal Travel Assistant", *In Autonomous Agents 98, Minneapolis MN USA*

[80] **Wooldridge, M. J. Jennings N. R.** "Intelligent Agents: Theory and Practice", *The Knowledge Engineering Review 10(2), 1995*